\documentclass[10pt,leqno]{amsart}
\usepackage{graphicx}
\usepackage[rightcaption]{sidecap}
\usepackage{indentfirst,csquotes, caption}
\usepackage{todonotes}

\usepackage[
backend=biber,
style=apa,
sorting=ynt
]{biblatex}
\usepackage{array}
\usepackage{amssymb,amsthm,amsmath}
\usepackage{xcolor,paralist,hyperref,fancyhdr,etoolbox}

\begin{document}

\title{Dimensions: Calculating Disruption Indices at Scale} 
\author[M. Pasin]{Michele Pasin}
\author[J. Sixt]{Joerg Sixt}

\date{\today}
\address{Michele Pasin, Digital Science, London, UK, https://orcid.org/0000-0001-8909-7766}
\email{m.pasin@digital-science.com}
\address{Joerg Sixt, Digital Science, London, UK, https://orcid.org/0000-0002-3475-6426}
\email{j.sixt@digital-science.com}


\let\thefootnote\relax

\begin{abstract}
Evaluating the disruptive nature of academic ideas is a new area of research evaluation that moves beyond standard citation-based metrics by taking into account the  broader citation context of publications or patents. The "$CD$ index" and a number of related indicators have been proposed in order to characterise mathematically the disruptiveness of scientific publications or patents. This research area has generated a lot of attention in recent years, yet there is no general consensus on the significance and reliability of disruption indices. More experimentation and evaluation would be desirable, however is hampered by the fact that these indicators are expensive and time-consuming to calculate, especially if done at scale on large citation networks. We present a novel method to calculate disruption indices that leverages the Dimensions cloud-based research infrastructure and reduces the computational time taken to produce such indices by an order of magnitude, as well as making available such functionalities within an online environment that requires no set-up efforts. We explain the novel algorithm and describe how its results align with preexisting implementations of disruption indicators. This method will enable researchers to develop, validate and improve mathematical disruption models more quickly and with more precision, thus contributing to the development of this new research area.  
\end{abstract} 
\maketitle

\section{Introduction}
Evaluating the disruptive nature of academic ideas is a new and promising area of research evaluation that moves beyond standard citation-based metrics by taking into account the  broader citation context of publications or patents. The idea of characterising scientific innovation in terms of its ‘disruptive’ property dates back to the work of \cite{bib29} and \cite{bib28} in the sociology and philosophy of science. These authors drew a fundamental distinction between contributions that improve pre-established scientific theories, and hence \textit{consolidate} their status as accepted truths, versus contributions that propose new or alternative methods that break away from the tradition, thus \textit{disrupting} it. 

In recent years, researchers working in the scientometrics and science of science (\cite{bib32}) communities have been proposing quantitative approaches for identifying disruptiveness. Identifying or predicting disruptive scientific ideas allows to understand the significance of scientists’ work in novel ways, as disruptive ideas not just impact the trajectory of scientific research, but also contribute to rendering obsolete the science that predates it. Two technological advancements underlie these developments: firstly, the growth of large, programmatically accessible bibliometric databases such as those provided by Dimensions, Crossref, Scopus and Web of Science (\cite{bib30}, \cite{bib31}); secondly, major advances in computing capabilities that facilitate the aggregation and processing of large-scale data sets, often using off-the-shelf infrastructure that requires minimal set up efforts for the scientometric researcher as it is available in the ‘cloud’ (\cite{bib4}). These two aspects combined permit to develop disruption metrics at scale, that is, by taking into account the not just a subset of scientific documents but the entire corpus of publications or patents as a single giant citations network. 

\cite{bib1} suggest a citation-based metric called "$CD$ index" to detect disruptive or consolidating publications or patents. The $CD$ index quantifies the degree to which future work cites a focal work together with its predecessors (that is, the references in the bibliography of the focal work). Disruptive papers are identified based on how much subsequent research cites them without citing their references (i.e. the papers they themselves cite). In essence, the $CD$ index tries to measure disruption by characterising the network of citations around a focal paper or patent. 

The $CD$ index has attracted particular attention ever since. In \cite{bib2} and \cite{bib15} the metric was used to analyse global trends in science. It has been used to detect disruptive publications in certain fields of clinical medicine (e.g. \cite{bib14}, \cite{bib17}, \cite{bib18}, \cite{bib19}, \cite{bib20} and \cite{bib21}) with various degrees of success. There have also been critics of the $CD$ index and its applications (e.g. \cite{bib24} and \cite{bib25}). Others have tried to test the metric's validity (e.g. \cite{bib7} and \cite{bib23}) in more general terms or create derivatives (e.g. \cite{bib26} and \cite{bib27}) with the aim to enhance its ability to pick truly game-changing science.

Despite increased interest in the $CD$ index e.g. due to \cite{bib2}, there are no easily accessible datasets and means of computation in order to assess its usefulness in the context of science of science analyses and literature discovery. One of the main challenges in investigating this new metric is the fact that its calculation is computationally expensive. This is a multi-dimensional metric \cite{bib6} as it involves not only direct linkages between nodes (i.e. citations of patents or publications) but also parts of their peripheral networks.

In this paper we present a method to compute the $CD$ index via a short SQL query on the Dimensions publication table on Google BigQuery (GBQ) (\cite{bib4},  \cite{bib5}). Our approach makes it possible to calculate e.g. $CD_5$ index for all journal articles with references in Dimensions in less than 5 hours.

Thanks to significantly quicker turnaround time, it is possible to test different variations of the $CD$ index and benchmark them against each other e.g. for other time spans or only using a subset of the Dimensions publication network like e.g. only looking at journal articles and excluding books, preprints, etc. This methodology also allows for more experimentation   e.g. it makes it easier to calculate the $CD$ index not for individual publications but per organization, or using it for patents which also exist as Dimensions GBQ tables.

\section{Method: calculating the \texorpdfstring{$CD$}{CD} index}

\subsection{The original definition of the \texorpdfstring{$CD$}{CD} index}
In its original definition from \cite{bib1} and subsequent articles \cite{bib2}, the $CD$ index is defined as an indicator for quantifying the degree to which future work cites a focal work together with its predecessors (that is, the references in the bibliography of the focal work). There has been a number of refinements and criticisms of this approach in recent years (e.g. \cite{bib6}, \cite{bib7}, \cite{bib8}) but in this article we focus primarily on the original definition from  \cite{bib1}, as the goal is to exemplify an improved implementation strategy that, we argue, can be easily adapted to suit any alternate $CD$ index formula.

\begin{figure}
    \centering
    \includegraphics[width=0.9\textwidth]{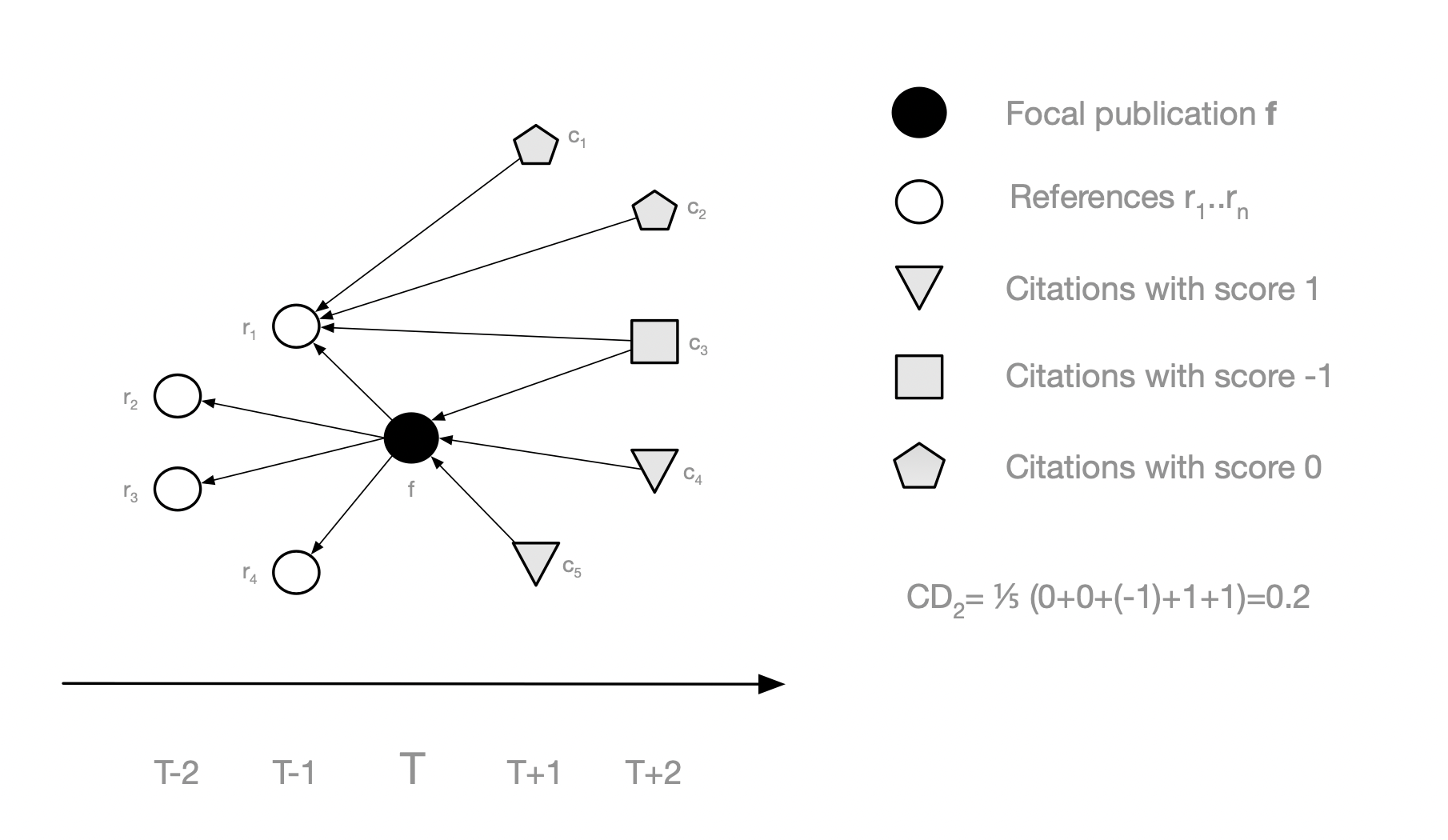}
    \caption{Example of a publication citation network around a focal publication. The x-axis is the timeline and indicates when the publications represented by squares, circles and pentagons are published. An arrow points from a citing publication to a publication it cites.}
    \label{fig1}
\end{figure}

The calculation of the index for a fixed focal publication relies on all its references and all citations to the focal publication and the references. Figure \ref{fig1} shows a simple example.

In its original form, the $CD$-index can be calculated as follows: 
\begin{enumerate}
    \item Fix a focal publication $f$ (the full circle in the diagram) published in year $T$ for which we want to calculate $CD_t$.
    \item Fix an integer $t$ that determines the time frame for which we want to measure impact: we will look at citations that occur at most $t$ years after the publication of $f$.
    \item Find all publications  $r_1\dots,r_k$ that are cited by $f$ (the empty circles in the diagram)  or in other words the "predecessors" or references of $f$.
    \item Find all $n$ distinct publications $c_1, \dots, c_n$ that cite at least one of the $f$, $r_1\dots,r_k$ in the years $T+1$ until and including $T+t$ (in other words the "successors" of $f$ or the union of all citations to $f$, $r_1\dots,r_k$ that occurred in the $t$ years after the publication of $f$)
    \item Assign a score $s(c_i)$ to each $c_i$ depending on what publication it cites:
    \begin{enumerate}
        \item Set $s(c_i):= 1$ if and only if $c_i$ cites $f$  but none of the references  $r_1\dots,r_k$ (the grey triangles in the diagram): the idea here is that such a citation does not care about the references but only about the focal paper $f$ highlighting the disruptive character of $f$.
        \item Set  $s(c_i):= -1$ if and only if $c_i$ cites at least one of the references  $r_1\dots,r_k$ and, in addition, also cites $f$ (the grey square in the diagram): the idea here is that such a citation cares about the references and the focal publication $f$ because $f$ consolidates the literature.
        \item Set  $s(c_i):= 0$ if and only if $c_i$ cites at least one of the references  $r_1\dots,r_k$ but does not cite $f$ (the grey pentagrams in the diagram): this means that $c_i$ covers similar topics as $f$ (after all it cites one or more references of $f$) but it ignores $f$ because $f$ is not significant.
    \end{enumerate}
    \item The $CD$-index is the average of all those scores i.e. 
    $$CD_t := \frac{1}{n}\sum_{i=1}^n s(c_i)$$
    Clearly $CD_t$ is a number between $-1$ and $1$.
\end{enumerate}

Accordingly, the two-year index $CD_2$ for the above diagram can be calculated as follows: the pentagon citations only cite the references and therefore receive a score 0, the triangles cite only $f$ and therefore receive score $1$ and the square cite both $f$ and its references and receive a score $-1$. All in all we have $5$ citations and therefore $CD_2= \frac{1}{5} (0+0+(-1)+1+1)=0.2$.
Note that $CD_1$ would only consider the $2$ citations taking place at $T+1$ and therefore $CD_1=\frac{1}{2} (0+1) = 0.5$. (This also illustrates that the parameter $t$ can have a significant effect on the index.)

\subsection{The challenge}
From a purely algorithmic perspective there are no issues with this method. It can be implemented in Python (\cite{bib3}) or other languages and run on datasets usually provided from third parties like Elsevier Scopus or Clarivate in the form of CSV files, etc. Calculations of the index for a few publications will be fast. However, anecdotal evidence suggests that calculations for a large set of publications can take many days.

An alternative approach is to store the publication and citation information in a database and run the calculation via SQL. Dimensions' publication data is already available as a GBQ table (\cite{bib5}) and can be queried in SQL. GBQ and SQL are very fast and can handle vast amounts of data. This led us to hope that this is a quicker way to calculate the index. The challenge here is the restrictive nature of SQL. Unlike Python, Java, etc. iterative routines and procedures are difficult to implement in SQL. Therefore the original algorithm needs to be translated into a different method compatible with SQL. The following sections explain this alternative way of calculation and the resulting SQL query.

\subsection{An alternative calculation method}

The original method requires us to first collect all citations $c_i$ to the focal paper $f$ and all its references $r_i$ and then in a next step check each of the $c_i$ again if they cite $f$ or not and if they cite one of the $r_i$ or not. In a sense we need to either go through all citations of the focal paper and its references twice or somehow remember where the citations have come from. We are not aware if this approach can be easily implemented in SQL.

In contrast we propose a different method that does not require cross-checking citations to $f$ and the citations to its references. Instead we run through all citations to $f$ and assign an intermediate score to each of them. In a next step we independently run through all citations to the focal paper's references and assign another intermediate score to each of them. Summing up these scores then gives us the final $CD$-index. As a result, this algorithm can be successfully expressed via SQL. 

\begin{figure}
    \centering
    \includegraphics[width=0.9\textwidth]{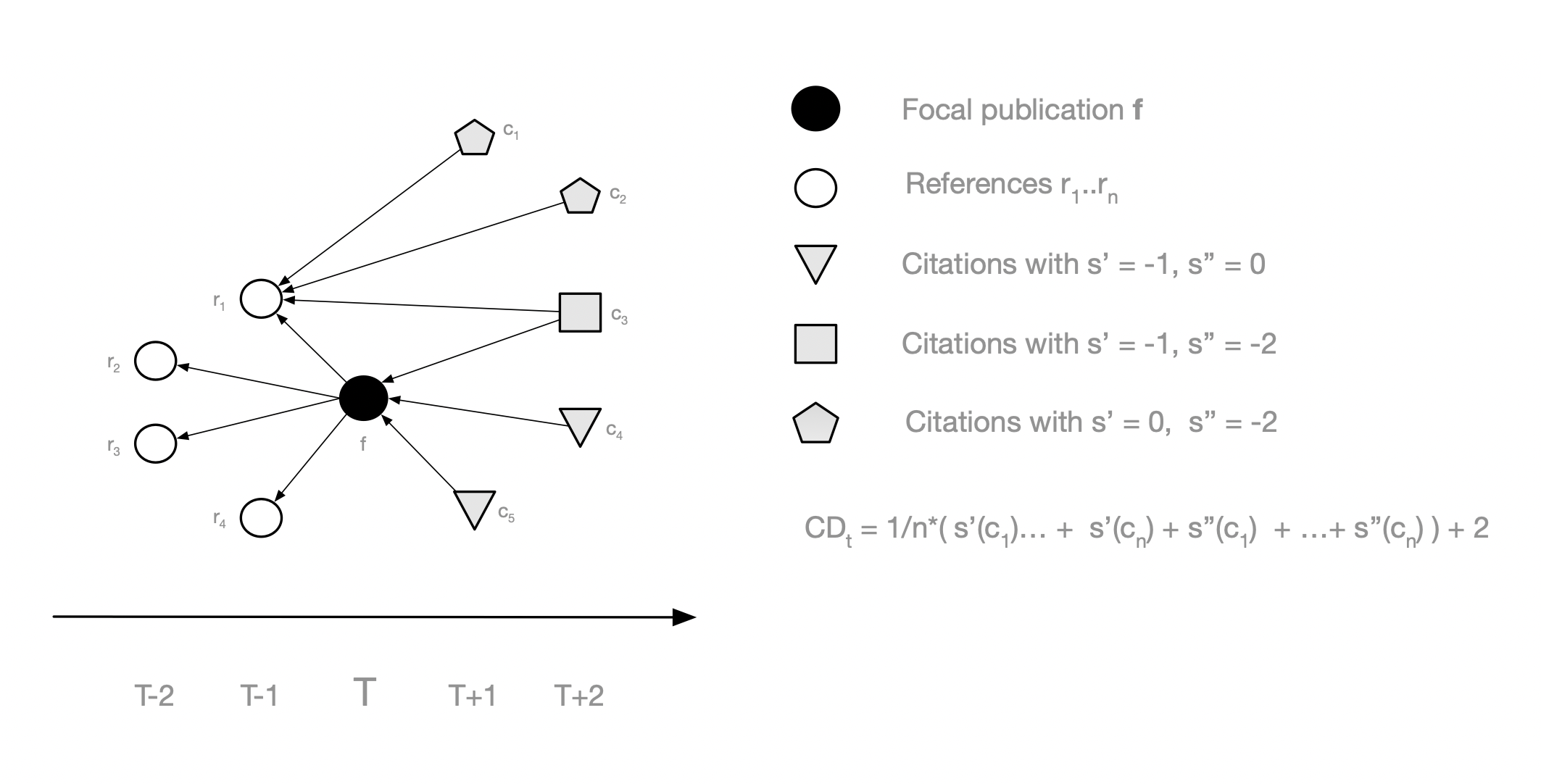}
    \caption{Example of a publication citation network around a focal publication}
    \label{fig2}
\end{figure}

We walk through this alternative algorithm step by step. See figure \ref{fig2} below for a visual summary of this approach. 

\begin{enumerate}
    \item Just like in the original method we fix a focal paper $f$ published in year $T$, an integer $T$, $f$'s references $r_1,\dots,r_k$ and the citations $c_1,\dots,c_n$ of any of the $f$, $r_1,\dots,r_k$ that occurred between $T+1$ and $T+t$.
    \item Assign each citation $c$ to $f$ (regardless if they cite any of the $r_1,\dots,r_k$ or not) a score $s'(c):=-1$ and $s'(x):=0$ for all other publications $x$.
    \item Assign each citation $c$ to any of the $r_1,\dots,r_k$ (regardless if they cite $f$ or not) a score $s''(c)=-2$ and $s''(x)=0$ for all other publications $x$.
    \item The $CD$-index is then  
    $$ CD_t = \frac{1}{n}\left(\sum_{i=1}^n s'(c_i) +  \sum_{i=1}^n s''(c_i)\right) + 2$$
\end{enumerate}

This method is more complex but will help us to create a SQL statement in the next section. Before we look at an implementation however we need to prove that both methods lead indeed to the same result. First of all observe that we can rewrite the formula as 
$$ CD_t = \frac{1}{n}\sum_{i=1}^n \left(s'(c_i) + s''(c_i) +2\right)$$
Therefore it is enough to show that $s(c) = s'(c)+ s''(c)+2$ for any $c$ in $\left\{c_1,\dots,c_n\right\}$. This can be easily verified by running through all the cases. Each $c$ in $\left\{c_1,\dots,c_n\right\}$ falls in exactly one of the following categories:
\begin{enumerate}
    \item If $c$ cites $f$ but none of the $r_1,\dots,r_k$. (i.e. $c$ is one of the grey square in the illustration) then $s'(c)+s''(c)+2 = (-1)+0+2=1$ which is exactly $s(c)$ from the original algorithm
    \item  If $c$ does not cite $f$ but it cites at least one of the $r_1,\dots,r_k$  (i.e. $c$ is one of the the grey pentagons in the illustration) then $s'(c)+s''(c)+2 = 0+(-2)+2=0$ which is exactly $s(c)$ from the original algorithm.
    \item If $c$ cites $f$ and also cites at least one of the $r_1,\dots,r_k$ (i.e. $c$ is one of the the empty squares in the illustration) then $s'(c)+s''(c)+2 = (-1)+(-2)+2=-1$ which is exactly $s(c)$ from the original algorithm.
\end{enumerate}
Hence the two methods lead to the same result. 

\subsection{The SQL statement}

In this section we will translate the alternative algorithm into SQL. The starting point is a table with a row for each publication with the following fields:
\begin{enumerate}
    \item \textbf{Publication ID}: a unique identifier for this publication e.g. DOI, PubMed ID or Dimensions publication ID
    \item \textbf{Publication year}: the year the publication has been published
    \item \textbf{Citations}: an array of all unique citations to this publication where each entry is a pair of a publication ID and citation (i.e. publication) year.
    \item \textbf{References}: an array of all unique publication IDs cited by this publication 
\end{enumerate}

Each of the IDs in citations and references needs to be an ID that is also included in the table. In the Dimensions publications GBQ table  \verb|dimensions-ai.data_analytics.publications| this data is already structured in that way with the fields \verb|id|, \verb|year|, \verb|citations| and \verb|reference_ids| (see \cite{bib5}).

The listing below is a simplified SQL statement that calculates $CD_5$ for \cite{bib9}.

\medskip

\begin{verbatim}
-- This is the focal publication f
DECLARE focal_publication_id STRING DEFAULT "pub.1019844293";
-- This is the impact span t
DECLARE time_diff INT64 DEFAULT 5;


WITH cd_raw_data AS
(
	-- Calculating s' for each citation to the focal publication
	-- All are assigned a score s'=-1. Any other publications appearing in
	-- the second SELECT and aren't included here
	-- implicitly get a score s'= 0
(
SELECT 
DISTINCT -- make sure we list unique citations otherwise we may double count
publications.id AS focal_id, -- focal publication
citation.id AS citation_id, -- citing publication to focal publication
-1 AS score -- s'
-- the Dimensions GBQ table for publications 
FROM `dimensions-ai.data_analytics.publications` AS publications
-- fetch all its citing publications: id and year
LEFT JOIN UNNEST(publications.citations) AS citation
-- for this experiment we only look at one publication
WHERE publications.id = focal_publication_id
-- we only consider citations that appear at most time_diff years after
-- the focal publication has been published
AND citation.year - publications.year BETWEEN 1 AND time_diff
)
UNION ALL
-- Calculating s'' for each citation to the references of 
-- the focal publication
	-- All are assigned a score s''=-2. Any other publications appearing in
	-- the first SELECT and aren't included here
	-- implicitly get a score s''= 0
(
SELECT DISTINCT 
publications.id as focal_id, -- focal publication
reference_citation.id as citation_id,-- citing publication to references
-2 as score -- s''
FROM `dimensions-ai.data_analytics.publications` as publications
-- get all the reference publication IDs of the focal publication
LEFT JOIN UNNEST(publications.reference_ids) as reference_id
-- get the references' meta data - mainly citations to it
INNER JOIN `dimensions-ai.data_analytics.publications` as references
ON references.id = reference_id
-- get the citations to the references
LEFT JOIN UNNEST(references.citations) as reference_citation
WHERE publications.id = focal_publication_id 
AND reference_citation.year - publications.year BETWEEN 1 AND time_diff
)
)
-- Now add up all scores, count the distinct ids of the citations in both SELECTs
-- above and use that information to calculate the $CD$-index
SELECT focal_id,
((SUM(score)/COUNT(DISTINCT citation_id))+2) as cd_index
FROM cd_raw_data
GROUP BY focal_id
\end{verbatim}

\medskip

At time of calculation the result was $-0.44$ which is not so far away from $-0.55$ listed in \cite{bib2} (which also uses a different indexing service's publications and citation data).

It is important to point out  that one issue with this method is that \verb|COUNT DISTINCT| in Google Big Query is a statistician function  and may not always be exact. For our purposes where we are looking at trends this is sufficient but if you need exact results for each and every publication you may need to use the computationally much more expensive \verb|EXACT_COUNT_DISTINCT| (see \cite{bib12}).

\section{Results}
\subsection{Calculating the \texorpdfstring{$CD$}{CD}-index for all publications}
\label{calccdsec}

The query that allows you to calculate the $CD$-index for all publications can be found in \cite{bib13}. Being able to access the Dimensions GBQ data is a prerequisite for running the query. Free of charge access for non-commercial scientometrics projects is available; also, it is possible to run the query on the freely available COVID dataset, although you will get different results.

We have run these queries in July 2023 to calculate $CD_5$ ($t=5$ is used most widely in the literature) for several citation networks:

\begin{enumerate}
    \item \textbf{All publications} (\verb|dim_all|): We computed the index for the complete list of 138m publications. Since not all publications have references and citations in the 5 year time frame the resulting table lists only 79m publications.
    \begin{enumerate}
        \item Query: \url{https://github.com/digital-science/dimensions-gbq-lab/blob/master/archive/CD-index/CD_index_query1_all.sql}
        \item Citation network: all publications
        \item Run: 28 July 2023, 18:00:06 UTC+1
        \item Duration: 4 hr 24 min
        \item Bytes processed: 69.62 GB
        \item Number of rows/publications: 79,095,524
        \item Total logical bytes 2.36 GB
    \end{enumerate}
    
    \item \textbf{Journal articles} (\verb|dim_journals|): In order to make the results more compatible with the calculation in the literature and in order to avoid artefacts in the metadata we also ran the algorithm for only journal articles with some references: (i.e. type is article and the journal ID is not null) with at least 10 references. The restriction of references is important because the definition of the $CD$-index gives any publication with no references and at least one citation immediately an index of 1. However, lack of references is usually just a result of missing metadata for a publication.
    \begin{enumerate}
        \item Query:\url{https://github.com/digital-science/dimensions-gbq-lab/blob/master/archive/CD-index/CDindex_query2_journals.sql}
        \item Citation network: all publications with type = article, journal.id not null, at least 10 references
        \item Start: 28 July 2023, 13:36:48 UTC+1
        \item Duration: 3hr 58min
        \item Bytes processed: 72.26 GB
        \item Number of rows/publications: 38,612,179
        \item Total logical bytes: 1.15 GB
    \end{enumerate}
    
    \item \textbf{PubMed} (\verb|dim_pubmed|): For a later comparison we also run the calculation for all publications listed in PubMed.
    \begin{enumerate}
        \item Query: \url{https://github.com/digital-science/dimensions-gbq-lab/blob/master/archive/CD-index/CDindex_query3_pubmed.sql}
        \item Citation network: all publications with a pubmed ID
        \item Start: 29 Jul 2023, 07:52:46 UTC+1
        \item Duration:3 hr 4 min
        \item Bytes processed: 69.96 GB
        \item Number of rows: 28,165,474
        \item Total logical bytes: 859.54 MB
        \end{enumerate}
\end{enumerate}

Please note that it is in the nature of the $CD$-index that changing the underlying publication network will also change the resulting index.

Mathematically our results should be correct however implementation mistakes, etc. can happen and therefore we decided to validate our results. In the following we use different approaches to validate the data.

\subsection{Comparison with selected publications}
\cite{bib2} provides explicit calculations of 3 publications which are very similar to our results (see Table \ref{table1}). Note that \cite{bib2} uses data from Web of Science and PubMed whereas we use Dimensions data yet the results are in a similar range. We list our results for the two

\begin{table}[]
    \centering
    \begin{tabular}{|m{6em}|m{6em}|m{6em}|m{6em}|m{6em}|}
    \hline
Publication&
From \cite{bib2} &
\verb|dim_all| &
\verb|dim_journals| &
\verb|dim_pubmed|
\\
\hline
Baltimore 1970&
$-0.55$&
$-0.44$&
$-0.50$&
$-0.44$
\\
\hline
Kohn Sham 1965&
$-0.22$ &
$-0.26$&
$-0.29$&
NULL (not indexed by PubMed)
\\
\hline
Watson Crick 1953&
$0.52$&
$0.60$&
NULL (less than 5 references)&
$0.63$
\\
\hline
    \end{tabular}
    \caption{A comparison of $CD_5$ from \cite{bib2} and our calculations}
    \label{table1}
\end{table}

Since the values for the $CD$-index are closely concentrated around zero for most publications (see below) with values between $-1 $and $1$ the results are quite close.

\subsection{Comparison with Russel Funk's Python Library}
We created a small sub citation network based on Dimensions data for the above sample publications. We have fed the citation network both into the  \verb|cdindex| Python library and our SQL and we arrived at exactly the same numbers in both instances. The precise implementation can be found in \cite{bib13}.

\subsection{Comparison with calculations by Russel Funk}
We received two sample data sets with publication identifiers and DOIs and their $CD$-index calculated by Russel Funk. These are based on Web of Science data (1m publications) and PubMed (2.3m publications). Different indexing services will have different citation networks which will affect the $CD$-index:
\begin{enumerate}
    \item Type of publications considered will restrict to certain references and citations e.g. PubMed covers (bio)medical and life science literature
    \item Time of running the query: a calculation run at time $T$ compared to another calculation at $T+x$ will miss out on considering citations that happened between $T$ and $T+x$. Even older citations may suddenly appear or disappear e.g. if the indexing service improves data processing or if older publications are disqualified or additional older data sources get included.
    \item Different indexation services use different ways to extract references and citations which can lead to differences in how citations are recognised
\end{enumerate}

\begin{table}[]
    \centering
    \begin{tabular}{|m{6em}||m{6em}||m{6em}|m{3em}|m{3em}|m{3em}|m{3em}|m{3em}|m{3em}|}
    \hline
source &
count &
mean &
std &
$q_{25}$ &
$q_{50}$ &
$q_{75}$ &
$q_{95}$ &
$q_{99}$
\\
\hline
\verb|funk_pubmed|&
$2326769$ &
$-0.01$ &
$0.11$ &
$-0.02$ &
$-0.00$ &
$-0.00$ &
$0.04$ &
$0.44$ 
\\
\hline
\verb|funk_wos|&
$836576$ &
$0.01$ &
$0.12$ &
$-0.01$ &
$-0.00$ &
$0.00$ &
$0.02$ &
$1.00$ 
\\
\hline
\verb|dim_all|&
$79095505$ &
$0.17$ &
$0.38$ &
$-0.00$ &
$0.00$ &
$0.01$ &
$1.00$ &
$1.00$ 
\\
\hline
\verb|dim_journals|&
$38612176$ &
$0.00$ &
$0.08$ &
$-0.01$ &
$-0.00$ &
$0.00$ &
$0.03$ &
$0.29$ 
\\
\hline
\verb|dim_pubmed| &
$28165467$ &
$0.15$ &
$0.37$ &
$-0.00$ &
$-0.00$ &
$0.00$ &
$1.00$ &
$1.00$ 
\\
\hline
    \end{tabular}
    \caption{Some basic statistical information of the various data fields. The minimum and maximum for all computations is -1 and 1.}
    \label{table2}
\end{table}

In Table \ref{table2} we can see some basic statistical information:
Source describes where the information came from:
\begin{enumerate}
    \item \verb|funk_pubmed|: an example dataset based on PubMed provided by R. Funk with 2.3m publications
    \item \verb|funk_wos|: an example dataset based on Web of Science provided by R. Funk with ca. 1m publications (however a number of them have no $CD$-index). Both datasets only included PubMed ID, DOI and the calculated $CD$-indices but no citation or references.
    \item \verb|dim_all,dim_journals,dim_pubmed|: our calculations, as per section \ref{calccdsec} above
    \item \verb|Count|: number of publications with non-NULL $CD_5$ index
    \item $q_*$: quantiles
\end{enumerate}

\begin{figure}
    \centering
    \includegraphics[width=0.9\textwidth]{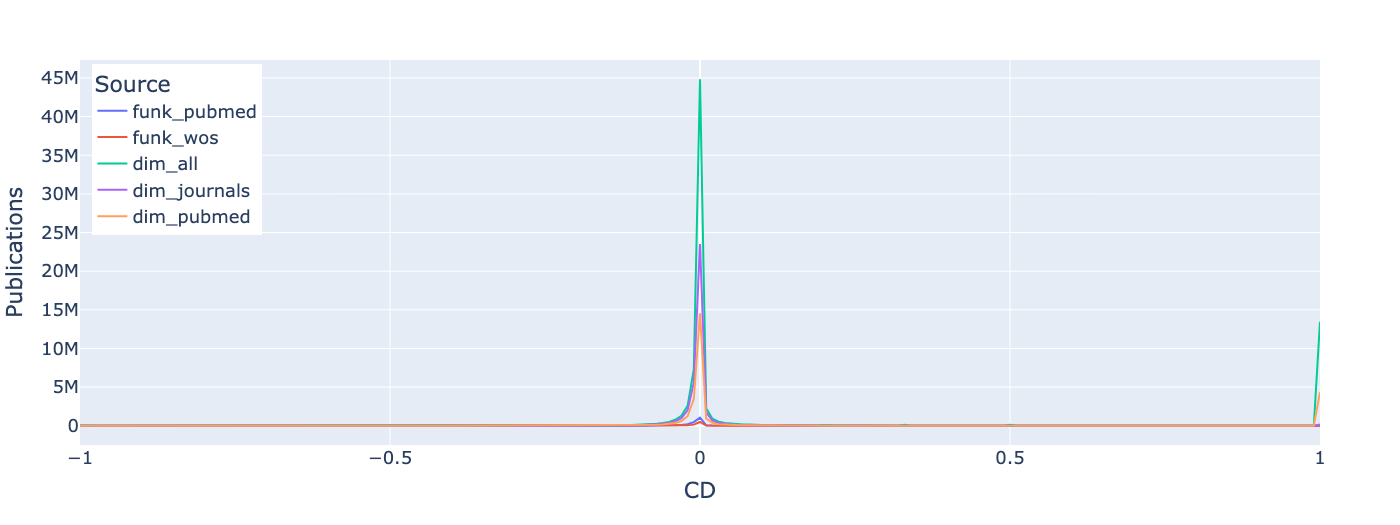}
    \caption{The distribution of the $CD_5$ index from the different sources. This is a histogram with bins of size $0.01$ for the $CD$-index. The data is also available in the file cd\_histogram.csv in \cite{bib13}}
    \label{fig4}
\end{figure}

We observe that all versions of the $CD$-index behave very similarly: a distribution around 0 which is concentrated around $0$. An exception is \verb|dim_all| which seem to have many more publications with a high $CD$-index. This is mainly due to the fact that there are 13m publications in the \verb|dim_all| dataset that have no references because Dimensions (or other services like CrossRef Dimension relies on) has not received the necessary metadata or full-text to extract references and citations. A simple histogram (Figure \ref{fig4}) of the 5 versions makes this even more evident.

\begin{figure}
    \centering
    \includegraphics[width=0.9\textwidth]{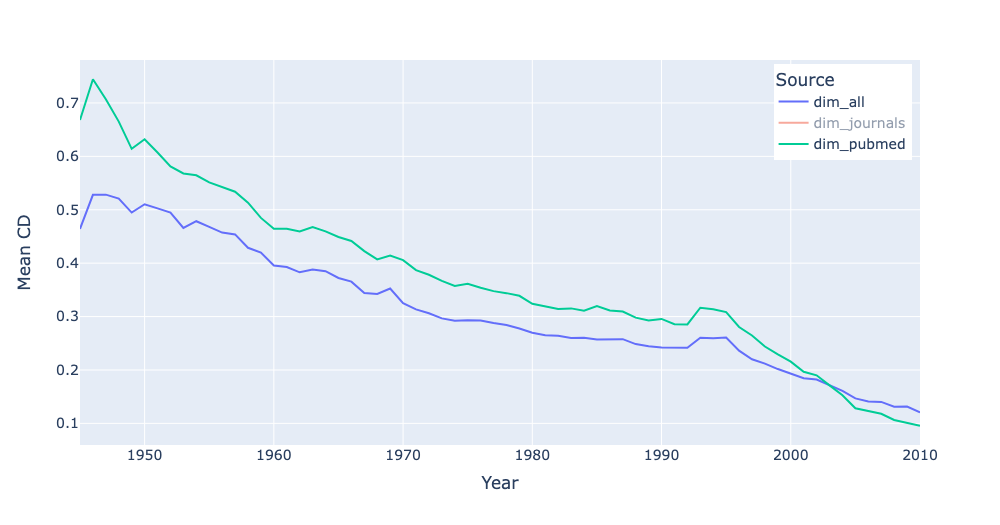}
    \caption{The average $CD_5$ index from our calculations over time. The data is also available in the file cd\_trends.csv in \cite{bib13}}
    \label{fig5}
\end{figure}

In Figure \ref{fig5} we also reproduced the decline of disruptive papers over time  visualised in \cite{bib1}.

At last we can also look at how much the sample data by R. Funk agrees with the one calculated by our method. As an example we use the PubMed data from R. Funk and Dimensions. These are 2.3m publications. We classify the top (bottom) 1\% for each $CD$-index as disruptive (consolidating) and the rest as neutral (following the terminology of \cite{bib1} where high $CD$-indices indicate disruption and low $CD$-indices consolidation). We considered if our methodology and the $CD$-index computation on PubMed data on Dimensions GBQ can mimic the Funk's $CD$-index created via PubMed data and his own calculations.

\begin{figure}
    \centering
    \includegraphics[width=0.9\textwidth]{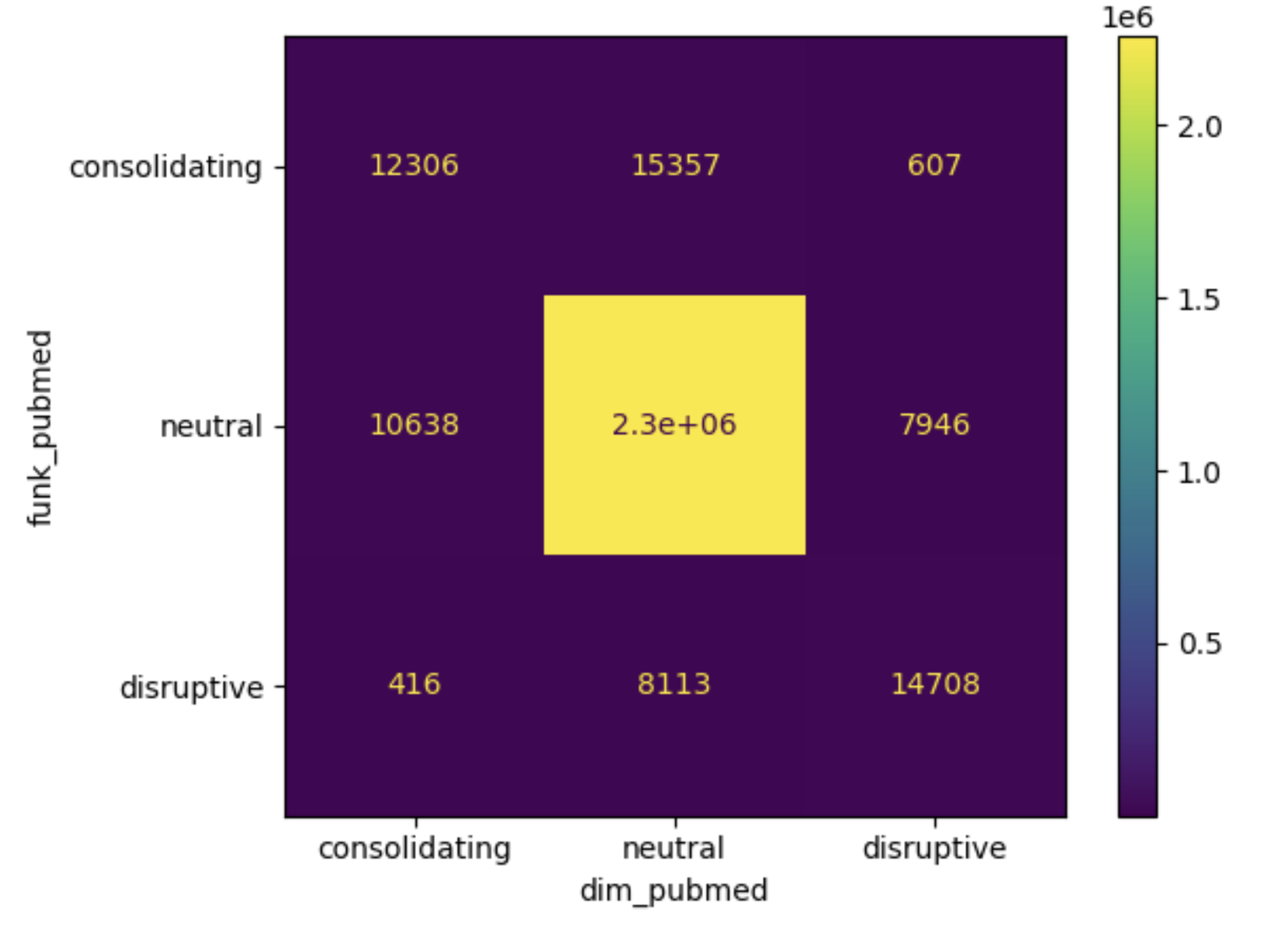}
    \caption{ The consolidation matrix for consolidating, neutral and disruptive publications according to Funk's PubMed calculations and our calculations.}
    \label{fig6}
\end{figure}

\begin{table}[]
    \centering
    \begin{tabular}{|m{6em}|m{5em}|m{5em}|m{5em}|m{5em}|}
    \hline
    &
precision&
recall&
f1-score&
support
\\\hline
consolidating &
$0.53$ &
$0.44$ &
$0.48$ &
$28270$ 
\\\hline
disruptive&
$0.63 $ &
$0.63$ &
$0.63$ &
$23237$ 
\\\hline
neutral&
$0.99$ &
$0.99$ &
$0.99 $ &
$2275205$ 
\\
\hline
    \end{tabular}
    \caption{Precision and recall if we use our calculations of $CD_5$ for PubMed publications to predict R. Funk's $CD_5$.}
    \label{table3}
\end{table}

Although precision and recall are not very impressive at least there are very few cases where one $CD$-index labels a publication as disruptive and the other $CD$-index labels it as consolidating and vice versa.

\section{Conclusion}
 In this article we presented a novel method for calculating disruption metrics based on SQL and the Dimensions on Google BigQuery data set, which reduces the computation time by an order of magnitude, when compared to traditional methods based on Python or other programming languages. Moreover, by leveraging the cloud-based architecture of Dimensions, this approach does not require specialised knowledge for setting up specialised computing infrastructure that can handle large-scale analytical tasks. Being able to calculate disruption metrics of publications and patents at scale, using multiple configurations and within reasonable amounts of time, makes it easier for researchers to focus on experimentation and analyses of these indicators, thus enabling the science of science community to assess and refine the usefulness of disruption indicators with increased confidence and speed. We validate our method against the original implementation of the CD index, both mathematically and by comparing the results of the calculations. The CD index results for the PubMed dataset and the code used to generate them is available online for review.

\bigskip
\subsubsection*{Conflicts of interest}
The authors are employees of Digital Science, the owner and commercial operator of Dimensions.

\subsubsection*{Acknowledgements}
We thank Russel Funk for providing us with some of his results for comparison with our own data and Daniel Hook for his advice.

\subsubsection*{Data Availability}
Jupyter notebooks and SQL queries for Dimensions on Google BigQuery are  available on Github \cite{bib13}
Dimensions on Google BigQuery data is available for non-commercial scientometrics research projects

\subsubsection*{Funding}
This research was not funded. The Open Access fees have been covered by Digital Science.

\subsubsection*{CRediT}
\begin{enumerate}
    \item Joerg Sixt: Conceptualization, Data curation, Formal Analysis, Investigation, Validation, Writing – original draft, Writing – review and editing
    \item Michele Pasin: Conceptualization, Project administration, Supervision, Validation, Writing – original draft, Writing – review and editing
\end{enumerate}

\medskip
\printbibliography

\end{document}